\documentclass{revtex4}
\usepackage{amsmath}
\usepackage{amssymb}
\usepackage{graphicx}

\begin{document}

\title{On the linear stability of spherically symmetric and wormhole solutions \\
supported by the sine-Gordon ghost scalar field}

\author{Vladimir Dzhunushaliev,$^{1,2,3}$
\footnote{
Email: vdzhunus@krsu.edu.kg}
Vladimir Folomeev,$^{2,3}$
\footnote{Email: vfolomeev@mail.ru}
Douglas Singleton$^{4}$
\footnote{Email: dougs@csufresno.edu}
and
Ratbay Myrzakulov$^{5}$
\footnote{Email: cnlpmyra1954@yahoo.com, cnlpmyra@mail.ru}
}
\affiliation{$^1$
Institute for Basic Research,
Eurasian National University,
Astana, 010008, Kazakhstan
\\ 
$^2$Institute of Physicotechnical Problems and Material Science of the NAS
of the
Kyrgyz Republic, 265 a, Chui Street, Bishkek, 720071,  Kyrgyz Republic \\
$^3$Institut f\"ur Physik, Universit\"at Oldenburg, Postfach 2503
D-26111 Oldenburg, Germany\\
$^{4}$ Physics Department, CSU Fresno, Fresno, CA 93740-8031\\
$^5$ Department of General and Theoretical Physics,
Eurasian National University, Astana, 010008, Kazakhstan\\
}

\begin{abstract}
In this paper we investigate wormhole and spherically symmetric solutions in 4D gravity plus a matter source consisting
of a ghost scalar field with a sine-Gordon potential. For the wormhole solutions we also include the possibility of electric and/or
magnetic charges. For both types of solutions we perform a linear stability analysis and show that the wormhole solutions are stable and that
when one turns on the electric and/or magnetic field the solution remains stable. The linear stability analysis of the spherically
symmetric solutions indicates that they can be stable or unstable depending on one of the parameters of the system. This result for the
spherically symmetric solution is nontrivial since a previous investigation of 4D gravity plus a ghost scalar field with a
$\lambda \phi ^4$ interaction found only unstable spherically symmetric solutions. Both the wormhole and spherically symmetric solutions
presented here asymptotically go to anti-de-Sitter space-time.
\end{abstract}

\maketitle

\section{Introduction}

The discovery of the accelerated expansion of the present Universe, has led to a search for a possible mechanisms of this
expansion. From the Einstein equations one finds that such an expansion is only
possible if one violates one of the energy conditions. In hydrodynamical language, this means that the
parameter of the equation of state  $w = \frac{p}{\rho} =\frac{pressure}{energy ~ density}$
should be less than $-1/3$ (i.e. violation of the strong energy condition), or
even less than $-1$ (i.e. violation of the weak energy or null energy condition). In this last case where
$w<-1$, one finds exponentially fast expansion
of the Universe. One of the simplest ways of obtaining such an acceleration is via
models with ghost scalar fields. In these models, one assumes that the sign in front of the
kinetic term of the scalar field is the opposite from that of the standard scalar field. It is well-known that ordinary
scalar fields only lead to an equation of state with $w\geq-1$. Ghost, scalar fields can give an equation of state with $w<-1$
in which case they are called phantom fields \footnote{In this work we use the term ghost scalar field -- a scalar field with the
opposite sign in front of the kinetic energy term. A phantom field is generically a field having $w<-1$. Although ghost fields can
lead to an equation of state with $w<-1$ the solutions found here have $w \ge -1$ for most or all of the range
of the radial coordinate, $r$, and for the parameters of the numerical solutions presented here. Thus we have a ghost scalar field but not a phantom field}.
Astronomical observations \cite{Tonry:2003zg,Alam:2003fg}, indicate that violation of the weak energy condition
is possible. Despite the strong theoretical misgivings about ghost fields these experimental observations have motivated
the theoretical consideration of such ghost scalar fields. Depending on the choice of the potential energy
of these ghost scalar fields one can obtain models of the accelerated expansion of the Universe
that are in good agreement with the observational data \cite{Elizalde:2004mq,Capozziello:2005tf}.

Ghost fields have also been studied in the context of compact astrophysical objects, such as stars and wormholes
composed partly or completely of ghost fields. In particular, ghost fields find a natural application to wormholes
since these objects are know to generically require violation of the weak (or null) energy condition \cite{Thorne:1988}.
In the papers \cite{Kodama:1978dw,Kodama:1979} a ghost scalar field with a Mexican hat potential
was investigated and it was found that this system only had regular, stable solutions for topologically non-trivial
(wormhole-like) geometry. The works \cite{kuhfittig,lxli,ArmendarizPicon:2002km,Sushkov:2002ef,lobo,sushkov} further investigated
traversable Lorentzian wormholes refining the conditions on the type of matter or field that would
lead to such space-times.  In the papers \cite{Nojiri:1998wb,Nojiri:1999pc} wormhole solutions induced by quantum
effects in GUTS were proposed. A general overview of the subject of Lorentzian wormholes and
violations of the various energy conditions can be found in the book by Visser \cite{Visser}.

There have also been papers devoted to non-wormhole like solutions which have a trivial
topology. The paper \cite{bronn2} gives a general classification of nonsingular, static
spherically symmetric solutions to the Einstein equations for  scalar fields with arbitrary
potentials and negative kinetic energy. In \cite{Dzh:2008} it was found that there existed spherically
symmetric solutions to the system of 4D gravity plus a ghost field with a Mexican hat potential which were
similar to the Bartnik-McKinnion solutions \cite{bk} but with the ghost field replacing the Yang-Mills field.
A linear stability analysis of the solutions found in \cite{Dzh:2008} indicated that these
solutions were not stable, although depending on the form of the effective potential they might
persist for long times -- i.e. they could be meta-stable.

Having a negative kinetic energy term leads to obvious quantum instability, but there are
possible solutions of this problem of quantum instability \cite{Nojiri,Carroll}. However, the question
of the classical, linear stability of a solution with ghost scalar fields requires a case by case study,
for example: (i) In \cite{Kodama:1979} it is shown that the solutions with the Mexican hat potential are stable
against radial perturbations. (ii) In \cite{Kim:2001ri} the classical, stability conditions for a charged wormhole created by a
massless, ghost scalar field were found. (iii) In \cite{bronn3,bronn4} it is shown that
wormhole solutions within the general framework of scalar-tensor theories with massless, non-minimally coupled scalar fields
are unstable under spherically symmetric perturbations. (iv) In \cite{Bronnikov:2002qx,Bronnikov:2005an} the generalization of this problem
was carried out for the case with electric or magnetic charges. (v) In \cite{ArmendarizPicon:2002km} a linear stability analysis for
a wormhole created by a massless ghost scalar field was carried out with the result that these wormholes had two asymptotically spherically symmetric ends
with either both ends having vanishing ADM masses or with the ends having ADM masses of opposite signs. It was found that such wormholes are stable
if the ADM masses are sufficiently small. (vi) In the paper \cite{Matos:2005uh} some estimates of the stability of a rotating, scalar field wormhole
were given. (vii) In \cite{Gonzalez:2008wd,Gonzalez:2008xk} it was shown that wormholes with a {\it massless} ghost scalar field are unstable with respect to linear
and nonlinear perturbations. (viii) A finally example of such classical, stability analysis can be found in \cite{Gonzalez:2009hn}
where it is shown that adding a charge to the model of (vii) did not lead to a stable wormhole solution.

These examples show that the question of classical stability or instability of a spherically symmetric or
wormhole configuration is strongly model-dependent and that one needs to carry out a stability analysis in each specific case.
In this paper, we carry out a linear stability analysis for a ghost scalar field model with the a sine-Gordon potential. We also
show why this system does not admit a topological or global stability analysis.
In a previous paper \cite{Dzhunushaliev:2009yw} this system was studied with the interesting result that one
could use the solutions to generate a dynamical model for the creation/annihilation of wormholes -- by allowing the
throat radius to vary smoothly from some finite value down to zero one could interpret this process as the
creation a wormhole with subsequent annihilation into two, disconnected spherically symmetric
solutions. In this paper, we will generalize the solutions of \cite{Dzhunushaliev:2009yw} to the include
electric and/or magnetic charges, and carry out a linear stability analysis for all these solutions.

For the solutions found in \cite{Dzhunushaliev:2009yw} it was shown that the static spherically
symmetric solutions with a ghost field plus sine-Gordon potential tended asymptotically to
anti-de-Sitter (AdS) spacetime. AdS space-times have generated much interest due to the discovery
of the AdS/CFT correspondence \cite{Maldacena:1997re,Gubser:1998bc,Witten:1998qj}
which implies that a gravitational theory in an AdS space-time is dual to a conformal field theory (CFT)
on the boundary of the space-time. The question of the stability of such space-times has already received
considerable attention \cite{Abbott:1981ff,Kleban:2004bv,Nayeri:2004aj}. In \cite{Abbott:1981ff}, in
the context of Einstein's theory plus a negative cosmological constant, the stability
of asymptotically AdS metrics against linear perturbations was demonstrated. In contrast reference \cite{Kleban:2004bv}
pointed out that in some cases instabilities may arise for asymptotically AdS spaces.
This implies the necessity of performing a stability analysis for solutions with asymptotically AdS space-times
on a case by case basis.

A second motivation  to study the stability of the solutions presented below is related to the
existence of the stable, static, sine-Gordon solitonic solutions for ordinary scalar field in (1+1) dimensions
without gravity \cite{rajaraman}. The sine-Gordon solutions are stable because of topological reasons.
In this paper we show that for some range of parameters adding gravity and changing the ordinary scalar field
to a ghost scalar field leads to classically, linear stable solutions in (3+1) dimensions. For other ranges of the
parameters the solutions are unstable. Thus including gravity and allowing ghost fields leads to linear stable
solutions which is in contrast to ordinary scalar fields in (3+1) which do not have stable solutions. However,
in contrast to the sine-Gordon soliton kink solution, the present solutions do not obtain their stability from
topological arguments.

\section{Field equations with electric and magnetic charges}

We consider a model of a gravitating ghost scalar field
in the presence of the electromagnetic field $F_{ik}$. The Lagrangian for this
system is \cite{Dzhunushaliev:2009yw}
\begin{equation}
\label{lagrangian}
  L =-\frac{R}{16\pi G}-
      \frac{1}{2}\partial_\mu \varphi \partial^\mu
        \varphi -V(\varphi)-\frac{1}{4}F_{lm}F^{lm}~,
\end{equation}
where $R$ is the scalar curvature, $G$ is Newton's gravitational constant,
and $V$ is the  sine-Gordon potential with a reversed sign from the usual case
\begin{equation}
\label{pot_mex2}
	V=\frac{m^4}{\lambda} \left[
		\cos\left(\frac{\sqrt{\lambda}}{m}\varphi\right)-1
	\right].
\end{equation}
Here $m$ is a mass of the field, and $\lambda$ is a coupling constant. The corresponding energy-momentum tensor is
\begin{equation}
\label{emt}
    T^k_i=
    -\partial_i \varphi \partial^k \varphi-F_i^l F^k_l-
        \delta^k_i \left[
            -\frac{1}{2}\partial_\mu \varphi \partial^\mu
            \varphi-V(\varphi)-\frac{1}{4}F_{lm}F^{lm}
        \right]~.
\end{equation}

To perform a linear stability analysis for this system, we take a time dependent, spherically symmetric  metric of the form
\begin{equation}
\label{metric_wh}
ds^2=e^{2 F(r,t)}dt^2-\frac{dr^2}{A(r,t)}-(r^2+r_0^2)(d\theta^2+\sin^2\theta d\phi^2),
\end{equation}
where the metric functions $F(r,t)$ and $A(r,t)$ depend both on the radial coordinate, $r$, and time, $t$.

Let us consider only radial components of electric and magnetic fields
$F_{01}=E_r$ and $F_{23}=-H_r$. Then by choosing the ansatz for the magnetic field
 $F_{23}=\partial_{\theta}A_3$, where $A_3=Q \cos\theta$ and $Q$ is the magnetic
charge, we have the following expression for the quadratic field combination
\begin{equation}
\label{invar}
F_{lm}F^{lm}=-2A e^{-2F}E_r^2+2\frac{Q^2}{(r^2+r_0^2)^2}.
\end{equation}
The equation for $E_r$ can be found from Maxwell's equations
\begin{equation}
\label{Maxw}
\left[\sqrt{-g}F^{ik}\right]_{,k}=0 \quad \Rightarrow \quad E_r=\frac{q e^{F}}{(r^2+r_0^2)\sqrt{A}},
\end{equation}
where $q$ is the electric charge.

Using the Lagrangian \eqref{lagrangian} and equations \eqref{invar} and \eqref{Maxw},
the gravitational equations take the form
\begin{eqnarray}
\label{ein_wh_sine}
	-\frac{A^\prime}{A}x+\frac{1}{A}+\frac{x^2}{x^2+x_0^2}-2&=&
	\beta\frac{x^2+x_0^2}{A}\left(-\frac{1}{2}e^{-2F}\dot\phi^2-\frac{A}{2}\phi^{\prime 2}
	+\cos\phi-1+\frac{\bar{q}^2+\bar{Q}^2}{2(x^2+x_0^2)^2}\right),
\\
\label{ein_wh_sine_11}
\frac{2 x^2}{x^2+x_0^2}-2-\frac{A^{\prime}}{A}x+2 x F^{\prime}&=&-\beta\frac{x^2+x_0^2}{A}\left[e^{-2F}\dot\phi^2+A\phi^{\prime 2}\right].
\end{eqnarray}
We have introduced new dimensionless variables $\phi=(\sqrt{\lambda}/m)\varphi,\, x=m r, \tau=m t$ and $\beta=8\pi G m^2/\lambda$,
$\bar{q}=\sqrt{\lambda}q, \bar{Q}=\sqrt{\lambda}Q$. Equations \eqref{ein_wh_sine} and \eqref{ein_wh_sine_11} are
the $(_t^t)$ and $\left[(_t^t)-(_x^x)\right]$ components of the Einstein equations.

The scalar field equation is
\begin{equation}
e^{-2F}\left[\ddot\phi-\left(\dot{F}+\frac{1}{2}\frac{\dot{A}}{A}\right)\dot\phi\right]-
	A\left[\phi^{\prime\prime}+\left(F^{\prime}+\frac{2x}{x^2+x_0^2}+\frac{1}{2}\frac{A^\prime}{A}\right)\phi^\prime\right]=
	-\sin\phi.
\label{field_wh_sine_0}
\end{equation}
In the above equations, ``prime" and ``dot" denote differentiation with respect to $x$ and $\tau$, respectively.

In \cite{Dzhunushaliev:2009yw} a set of static solutions was found for the system \eqref{ein_wh_sine}-\eqref{field_wh_sine_0}
for various values of the parameter $x_0$ and with the electric and magnetic charges ($\bar{q}, \bar{Q}$) set to zero. It was shown that
there exist regular solutions up to the value $x_0=0$. This was interpreted as the possible creation/annihilation process
of a wormhole -- as the throat radius  $x_0 \rightarrow 0$ the wormhole split into two, separate, spherically symmetric
space-times. This interpretation assumed that the static solutions could be considered as snapshots of a dynamical process where the
wormhole annihilated into two separate space-times or run in reverse where two separate space-times merged to
create a wormhole. These solutions and this creation/annihilation interpretation are similar to the 5D Kaluza-Klein dyonic wormholes
studied in \cite{Dzh:1998}

\begin{figure}[t]
\begin{minipage}[t]{.49\linewidth}
  \begin{center}
  \includegraphics[width=9.5cm]{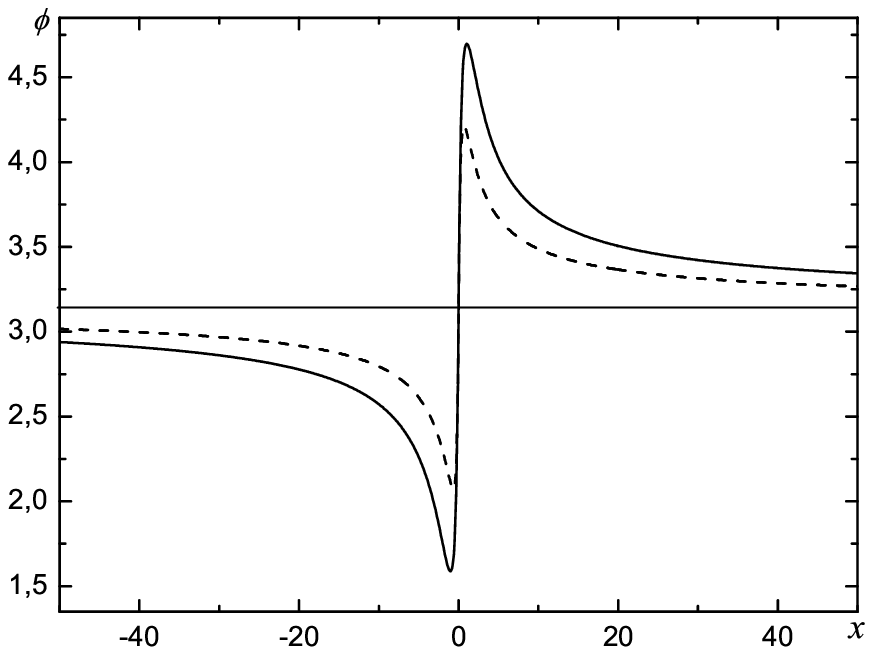}
\vspace{-1.cm}
  \caption{The background, static, wormhole solutions of the system \eqref{ein_wh_sine}-\eqref{field_wh_sine_0} for
 the scalar field $\phi$  at $x_0=0.3$, $\beta=1$ and
 $\bar{q}=0$ (the solid line),  $\bar{q}=0.35$ (the dashed line), $\bar{Q}=0$ for all graphs. Asymptotically $\phi\rightarrow \pi$.}
    \label{static_sols_phi}
  \end{center}
\end{minipage}\hfill
\begin{minipage}[t]{.49\linewidth}
  \begin{center}
  \includegraphics[width=9.7cm]{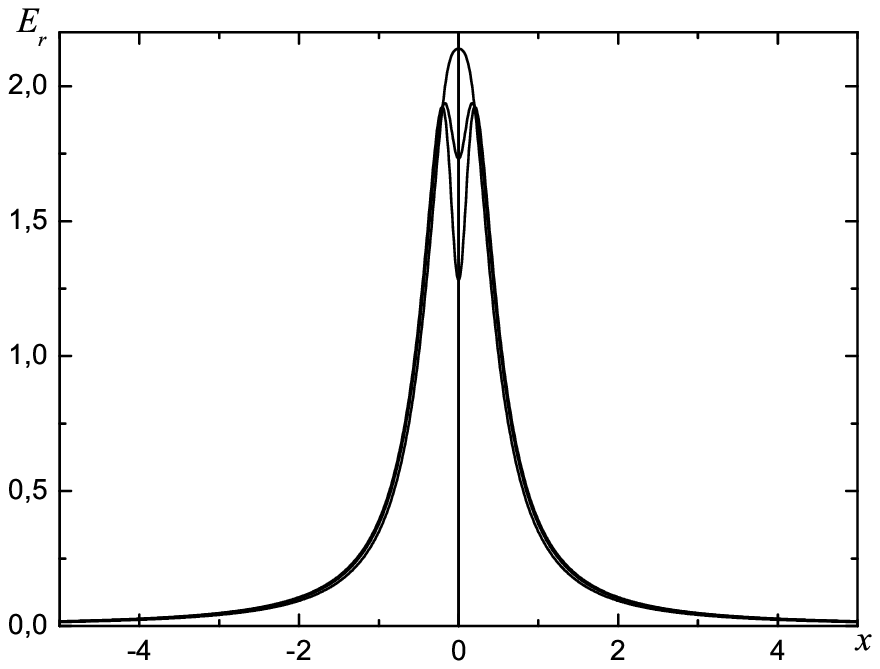}
\vspace{-1.1cm}
  \caption{The background, static, wormhole solutions of the system \eqref{ein_wh_sine}-\eqref{field_wh_sine_0} for
  the dimensionless electric field $\bar{E}_r=(\sqrt{\lambda}/m^2) E_r=\bar{q}e^F/\left[(x^2+x_0^2)\sqrt{A}\right]$
 at $x_0=0.3$, $\beta=1$ and $\bar{q}=0.35$,  $\bar{q}=0.40$, and  $\bar{q}=0.43$, top-down, respectively, $\bar{Q}=0$
 for all graphs. Asymptotically $\bar{E}\to \bar{q}/x^2$.}
\label{static_sols_Er}
  \end{center}
\end{minipage}\hfill
\end{figure}

\begin{figure}[h]
\begin{center}
  \includegraphics[width=11cm]{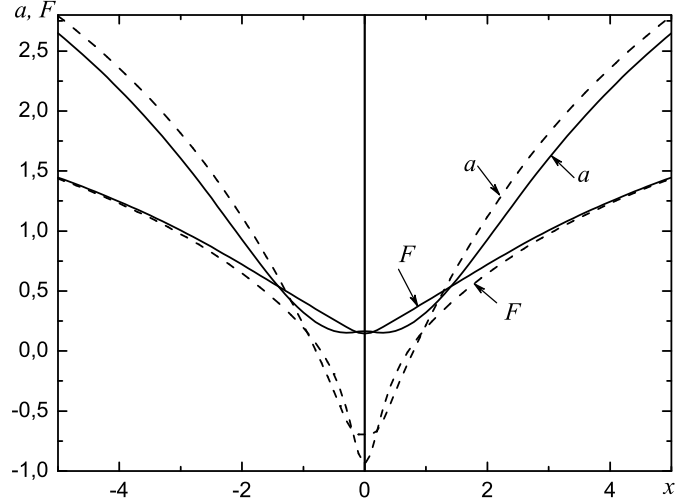}
\vspace{-1cm}
 \caption{The background static, wormhole solutions of the system \eqref{ein_wh_sine}-\eqref{field_wh_sine_0} for
 the metric functions $F$ and $a=\ln{A}$ at $x_0=0.3$, $\beta=1$ and
 $\bar{q}=0$ (the solid lines),  $\bar{q}=0.35$ (the dashed lines), $\bar{Q}=0$ for all graphs. Asymptotically
 $e^{2F}\rightarrow \frac{2\beta}{3}x^2$ and $A\rightarrow \frac{2\beta}{3}x^2$. This
 means that the space-time for large $x$ is Anti-de-Sitter.}
\label{static_sols_2}
\end{center}
\end{figure}

We now extend the results of \cite{Dzhunushaliev:2009yw} by including radial electric and magnetic fields.
These fields will give positive contributions to the total energy density of the system,
leading to solutions differing from the ones discussed in \cite{Dzhunushaliev:2009yw}. We have not found an analytical solution
for the system \eqref{ein_wh_sine}-\eqref{field_wh_sine_0} for the entire range of $x$. Thus we solve this system
of coupled, non-linear equations numerically. From the equations \eqref{ein_wh_sine}-\eqref{field_wh_sine_0} we obtain
the initial conditions for starting our numerical solution as
\begin{equation}
\label{ini_wh_sine}
F(0)={\rm const.},\quad	A(0)=1- \beta \left[\frac{\bar{q}^2 +\bar{Q}^2}{2 x_0^4}-2\right]x_0^2, \quad \phi(0)=\pi,
 \quad	\phi^\prime(0)=\sqrt{\frac{2}{\beta x_0^2}} ~.
\end{equation}
These initial conditions, where $x_0 \ne 0$, give wormhole solutions. Later we will discuss the spherically symmetric,
boson star solutions for which $x_0 = 0$.
From the above expression for $A(0)$ one can see that the addition of a nonzero electric and/or magnetic charges $\bar{q}, \bar{Q}$
gives a lower limit on  $x_0$ for which $A(0)$ is positive. We do not give the explicit expression for $x_0$ but it involves solving
a quadratic equation for $x_0 ^2$. The physical meaning is that the addition of a large charge -- and its associated positive energy density --
makes the existence of wormholes and spherically symmetric solutions impossible. The general, qualitative behavior of the solutions at
acceptable values of $x_0$ (i.e. values for which $A(0) >0$) with electric and/or magnetic fields
remains similar to the case when there are no fields (compare figures \ref{static_sols_phi}-\ref{static_sols_2} of the
present paper with  figures 1-3 in \cite{Dzhunushaliev:2009yw}). From now on we take into account only electric charge
by taking $\bar{q}$ non-zero and setting $\bar{Q} =0$. This is not a reduction in the generality of the system since,
from equation \eqref{ein_wh_sine}, it is possible to introduce a new effective electric charge defined as
$\bar{q}^2_{eff}=\bar{q}^2+\bar{Q}^2$. As implied by electromagnetic duality it is only a matter of convention
whether one talks of electric charge, magnetic charge or some combination. What matters is $\bar{q}_{eff}$. From the numerical
solutions shown in figures \ref{static_sols_phi}-\ref{static_sols_2} one can make the following statements about the asymptotic form
of the solutions: (i) The scalar field approaches the minimum of the potential $V(\varphi )$ from \eqref{pot_mex2} (i.e. $\phi = \pi$)
both from below (for $x<0$) and from above (for $x>0$). (ii) The electric/magnetic field asymptotically takes a Coulomb form,
$1/x^2$. Note as the effective electric/magnetic charge increases a cusp develops at $x=0$. This might be seen as the
influence of the electric/magnetic charge to generate a singularity at $x=0$ in the case of a space-time with
trivial topology i.e. when $x_0 \rightarrow 0$ and the wormhole space-time goes over to the spherically symmetric
topology. (iii) The metric ansatz functions $F(x), A(x)$ asymptotically go as
$e^{2F}\rightarrow \frac{2\beta}{3}x^2$ and $A\rightarrow \frac{2\beta}{3}x^2$ i.e.
asymptotically the metric approaches Anti-de-Sitter space-time. In order to have the space-time be asymptotically AdS one
needs to choose the constant value of $F(0)={\rm const.}$ such that both $e^{2F(x)}$ and  $A (x)$ approach
$\frac{2\beta}{3}x^2$ as $x \rightarrow \infty$. In figure \ref{static_sols_2}, for example, $F(0) \approx -0.945$
at $\bar{q}=0.35$ and $F(0) \approx 0.145$ at $\bar{q}=0$ in order to meet this requirement.
The choice of $F(0)$ is connected with the scaling of the time coordinate at $x=0$. Thus one must
appropriately scale the time coordinate at $x=0$ in order to have both metric functions, $e^{2F(x)}$ and  $A (x)$, give
asymptotic AdS space-time.

\section{Linear stability analysis}

We now study the dynamical stability of the above static solutions under linear perturbations with a harmonic time dependence.
This type of stability analysis is essentially that used in the works \cite{jetzer, jetzer2} to study boson
stars. The present authors have used this type of stability analysis to study the issue of stability for the
spherical solutions for the system of a ghost scalar field with a Mexican hat potential coupled to gravity \cite{Dzh:2008}.
For simplicity, we introduce a new metric function $a (x)$ in place of $A (x)$ defined as $A (x) =e^{a(x)}$.
We perturb the solutions of the system \eqref{ein_wh_sine}-\eqref{field_wh_sine_0} by expanding the metric functions and scalar
field function to first order as follows
\begin{eqnarray}
F(x,\tau)&=&F_0(x)+F_1(x)\cos\omega\tau, \nonumber\\
a(x,\tau)&=&a_0(x)+a_1(x)\cos\omega\tau, \nonumber\\
\phi(x,\tau)&=&\phi_0(x)+\phi_1(x)\frac{\cos\omega\tau}{\sqrt{x^2+x_0^2}} \nonumber.
\end{eqnarray}
The index 0 indicates the static background solutions of equations \eqref{ein_wh_sine}-\eqref{field_wh_sine_0}
and $\tau$ is the scaled time parameter defined in the previous section. The first-order perturbation
equations which follow from \eqref{ein_wh_sine} and \eqref{ein_wh_sine_11} are
\begin{eqnarray}
\label{metr_pert}
a_1^{\prime}&=&-\beta \frac{\sqrt{x^2+x_0^2}}{x}\left[\frac{x\phi_0^{\prime}\phi_1}{x^2+x_0^2}-
\phi_0^{\prime}\phi_1^{\prime}-e^{-a_0}\sin (\phi_0)\, \phi_1 \right],\\
F_1^{\prime}&=&\frac{1}{2}a_1^{\prime}-\beta\frac{\sqrt{x^2+x_0^2}}{x}\left[\phi_1^{\prime}-
\frac{x}{x^2+x_0^2}\,\phi_1\right]\phi_0^{\prime},
\end{eqnarray}
and the equation for $\phi_1$ is
\begin{equation}
\label{sf_pert}
\phi_1^{\prime\prime}+\left[F_0^{\prime}+\frac{1}{2}a_0^{\prime}\right]\phi_1^{\prime}-
V_0(x)\phi_1+\omega^2 e^{-2F_0-a_0}\phi_1=0
\end{equation}
with the potential given by
\begin{equation}
\label{pot_pert}
V_0(x)=e^{-a_0}\left[\cos\phi_0-\beta\frac{x^2+x_0^2}{x}\sin(\phi_0)\,\phi_0^{\prime}\right]-
\frac{x^2}{(x^2+x_0^2)^2}+\frac{1}{x^2+x_0^2}+\left(F_0^{\prime}+\frac{1}{2}a_0^{\prime}\right)\frac{x}{x^2+x_0^2}.
\end{equation}

\begin{figure}[t]
\begin{center}
  \includegraphics[width=14cm]{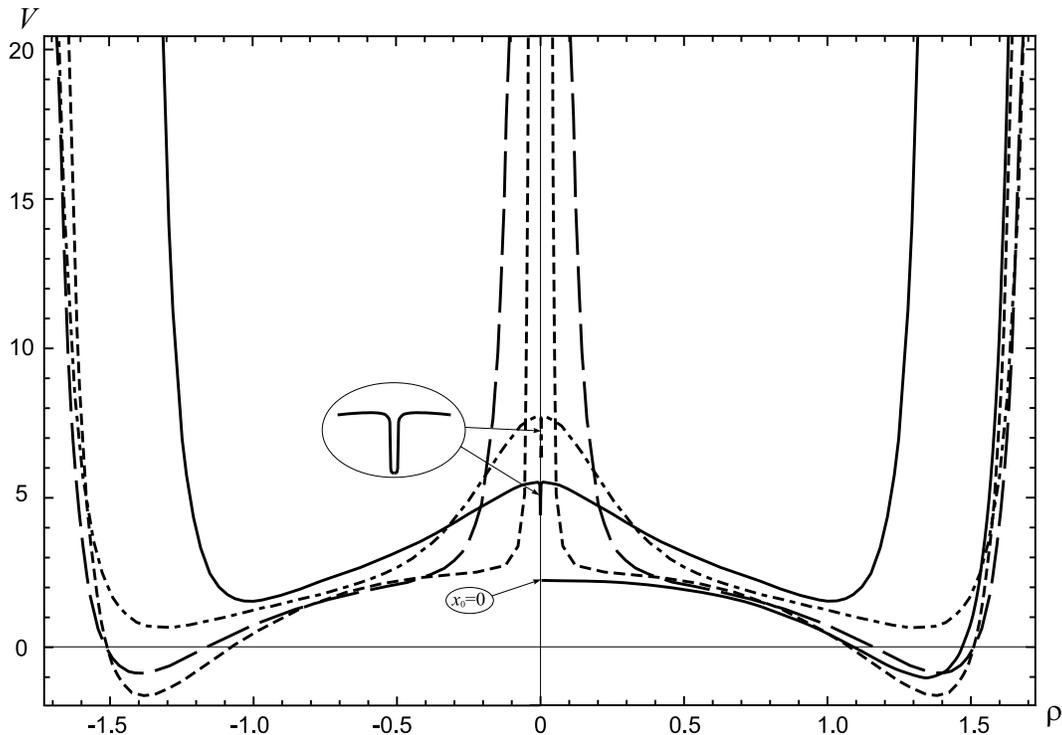}
 \caption{The potential $V[x(\rho)]$ from \eqref{pot_pert_2} as a function of $\rho$ which is defined in \eqref{new_indep}.
 The solid line corresponds to $x_0=1$, the dash-dotted, the long dashed and the short dashed lines correspond to $x_0=0.5$,
 $x_0=0.1$ and $x_0=0.01$, respectively. The curve for $x_0=0$ (the spherically symmetric solution) is also shown by the solid line.
 For the spherically symmetric solution the range of $\rho $ is $0 \le \rho \le \infty$.
 For all solutions the electric and magnetic charges are set to zero ($\bar{q},\bar{Q}=0$) and $\beta=1$. }
\label{stab_q_0}
\end{center}
\end{figure}

Introducing the new independent variable $\rho$
\begin{equation}
\label{new_indep}
\frac{d\rho}{d x}=\exp{\left(-F_0-\frac{1}{2}a_0\right)}
\end{equation}
we can rewrite equation \eqref{sf_pert} in a Schr\"{o}dinger-like form
\begin{equation}
\label{sf_pert_2}
-\frac{d^2\phi_1}{d\rho^2}+V[x(\rho)]\phi_1=\omega^2\phi_1,
\end{equation}
where
\begin{equation}
\label{pot_pert_2}
V[x(\rho)]=e^{2 F_0+a_0}V_0(x).
\end{equation}
The study of the stability of the static solutions displayed in figures
\ref{static_sols_phi}-\ref{static_sols_2} now reduces to a study of the Schr\"{o}dinger-like
equation given by \eqref{sf_pert_2} and \eqref{pot_pert_2}. If the energy-like
eigenvalues, $\omega ^2$ from \eqref{sf_pert_2} are $>0$ then the solution is stable; if
there are eigenvalues such that $\omega ^2 <0$ then the perturbations grow exponentially
and the solution is unstable.

Before moving on to the details of the linear stability analysis of the wormhole and spherically symmetric cases
we discuss several other criteria for stability of solutions similar to those presented here.
There are several other stability criteria -- topological stability \cite{rajaraman} and  global stability  \cite{kusmartsev} --
which are more powerful and go beyond the linear stability analysis presented here. However the present solution does not
meet the conditions that allow us to apply these broader stability criteria.

The typical example of topological stability is given by the sine-Gordon kink
in the absence of gravity \cite{rajaraman}. The sine-Gordon kink is topologically stable since the
value of its scalar field takes two different values at $x=-\infty$ and $x=+\infty$. From figure \ref{static_sols_phi}
one can see that the wormhole solution  does not have this feature; the scalar field of the wormhole
solution approaches the same value at  $x=-\infty$ and $x=+\infty$ thus ruling out a topological argument its stability.

A global stability analysis method for systems of gravity coupled to a complex scalar field was given in
\cite{kusmartsev}. In this work the global stability was determined by studying the binding energy of the system
$$
B=M-mN
$$
where $M$ is that Tolman mass and $N$ is the particle number. The Tolman mass is defined by
$$
M = \int (2 T_0 ^0 - T_\mu ^\mu ) \sqrt{|g|} d^3 x ~.
$$
From \eqref{emt} $T_\mu^\mu=\partial _\mu \phi \partial ^\mu \phi +4 V (\phi)$. Combining this with
$T_0 ^0$, which also can be obtained from \eqref{emt} or read off of
\eqref{ein_wh_sine}, we find that $(2 T_0 ^0 - T_\mu ^\mu )$ in the above equation tends asymptotically
to a constant value equal to $-2 V_\infty=4$, where $V_\infty$ is the asymptotic value of the
potential. The asymptotic value of $\sqrt{|g|} \propto r^2$ and so the integral for $M$ above diverges --
as one would expect for a space-time which is asymptotically AdS.

Next, the particle number $N$ is the integral over all space of the time component, $j^0$, of the Noether current density
$j ^\mu =\frac{i}{2} \sqrt{|g|} g^{\mu \nu}[\phi^* \partial _\nu \phi - \phi \partial _\nu \phi ^* ]$
$$
N = \int j^0 d^3 x ~.
$$
Since our scalar field is not complex $j ^\mu =0$ and so $N=0$. Thus neither the wormhole nor spherically symmetric
solutions presented in section II can be analyzed in terms of topological stability or global stability due to the
asymptotic AdS nature of the space-time, the asymptotic behavior of the scalar field or the fact that the scalar field
is real rather than complex.

\subsection{Wormhole case ($x_0 \neq 0$)}

We now study the stability of the wormhole solutions for various values of the wormhole throat size
$x_0$. Figure \ref{stab_q_0} shows four effective potential curves for $x_0=1, x_0=0.5, x_0=0.1$ and $x_0=0.01$ respectively and with no
electromagnetic field. One can see that for $x_0=1$ and $x_0=0.5$ the curves of the potential are completely
positive so that $\omega ^2 >0$ for any solutions of \eqref{sf_pert_2} and \eqref{pot_pert_2}
i.e. the solutions are stable. In the cases when
$x_0=0.1$ and $x_0=0.01$ there are small regions where the potential \eqref{pot_pert_2} becomes negative. From
figure \ref{stab_q_0} these negative regions occur around $\rho \approx 1.4$. Thus
in principle the lowest eigenvalue $\omega^2$ could be negative. Further analytical details which support the
positiveness of $\omega ^2$ for the parameters used in this paper are given in appendix A, even in
those cases when the potential $V[x (\rho)]$ has regions where it dips below zero. In the language of ordinary
quantum mechanics these regions are not deep enough for a bound state with $\omega ^2<0$ to form.

For comparison figure \ref{stab_q_0} also shows the behavior of the potential \eqref{pot_pert_2} for the
spherically symmetric solution, i.e., when $x_0=0$ and $\beta=1$. The full investigation of this case is given in
the next subsection and appendix B. One can see that near
$\rho=0$ behavior of the spherically symmetric case differs from the behavior of the wormhole solution as
$x_0 \to 0$. This difference is connected with different behavior of the derivative  $\phi^{\prime}(0)$
at  $x=0$: For the wormhole solution, this derivative tends to infinity, according to \eqref{ini_wh_sine}; for the
spherically symmetric solution, the derivative goes to $0$ \cite{Dzhunushaliev:2009yw}.
Nevertheless, at $\beta=1$, the spherically symmetric solution is stable, since the lowest eigenvalue, $\omega^2$, is positive.


\begin{figure}[t]
\begin{center}
  \includegraphics[width=9cm]{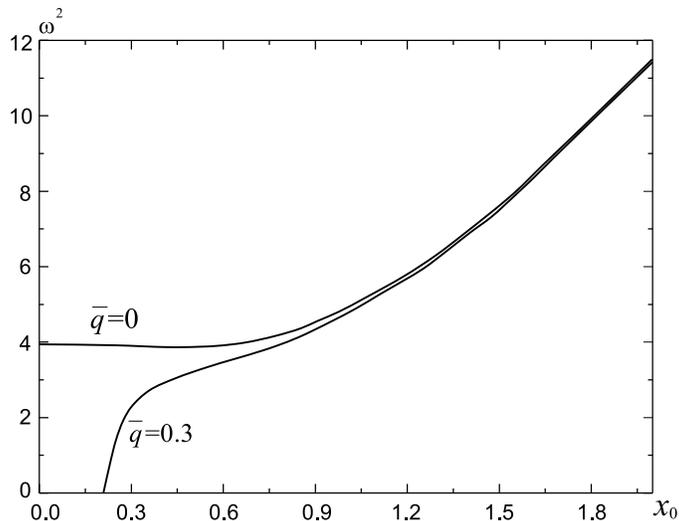}
 \caption{The dependence of the lowest eigenvalues $\omega^2$ on the throat size $x_0$ without electric charge ($\bar{q}=0$)
 and with $\bar{q}=0.3$, $\beta=1$.
 For $\bar{q}=0.3$ at $x_0\approx 0.21$, $\omega^2 \to 0$ and $A(0)\to 0$ simultaneously (see equation \eqref{ini_wh_sine}).}
\label{omega_x0}
\end{center}
\end{figure}

\subsection{Spherically symmetric case ($x_0=0$)}

\begin{figure}[t]
\begin{minipage}[t]{.49\linewidth}
  \begin{center}
  \includegraphics[width=9.5cm]{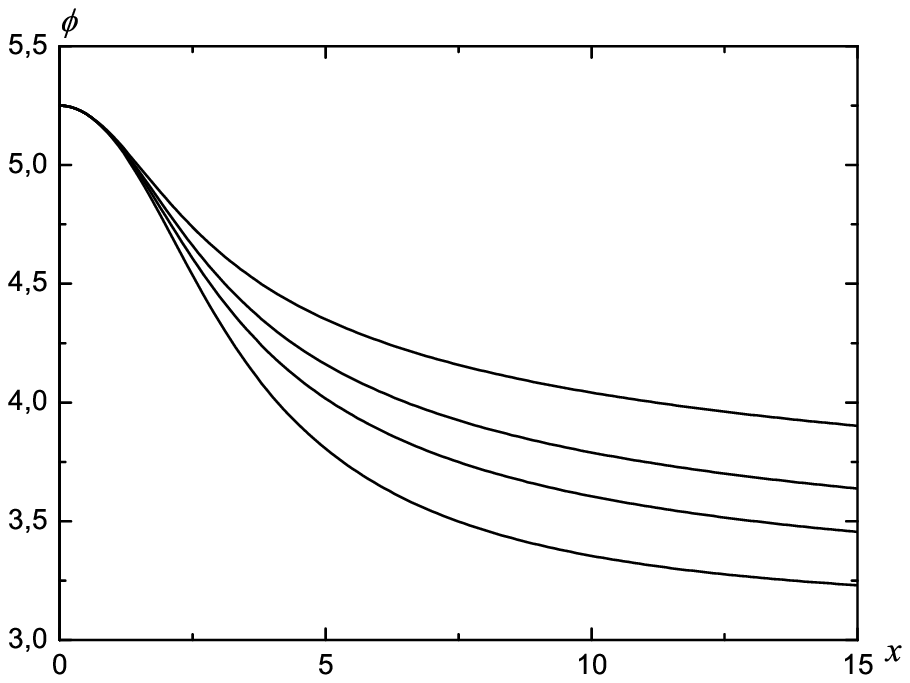}
\vspace{-1.5cm}
  \caption{The spherically symmetric case: The scalar field $\phi$ at different values of the parameter $\beta$: 1.5, 1.0, 0.75 and 0.5,
           from top to bottom. Asymptotically $\phi\rightarrow \pi$.}
    \label{phi_sphera}
  \end{center}
\end{minipage}\hfill
\begin{minipage}[t]{.49\linewidth}
  \begin{center}
  \includegraphics[width=9.5cm]{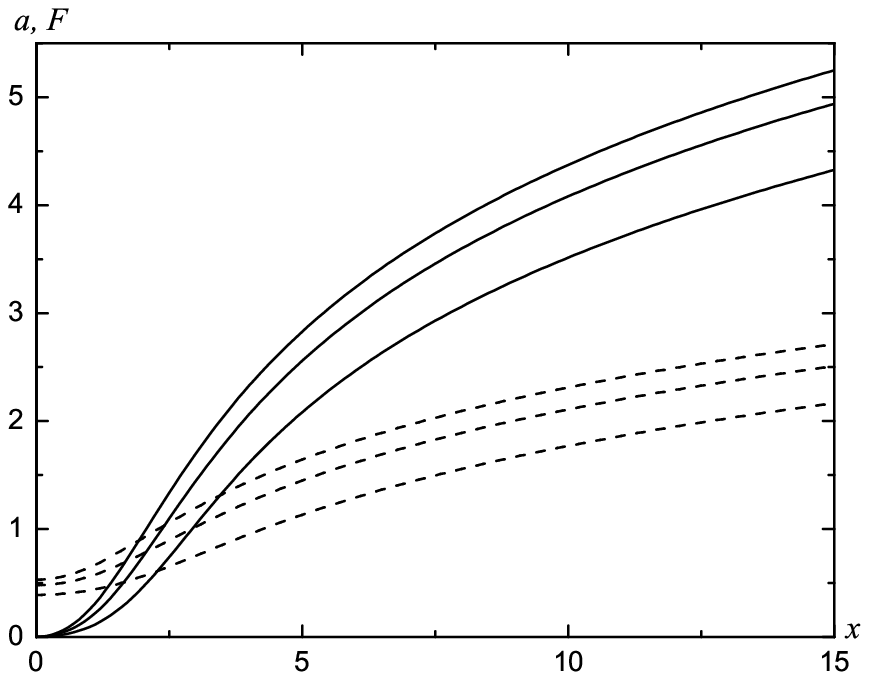}
\vspace{-1.5cm}
  \caption{The spherically symmetric case: The metric functions $F (x)$ (the dashed lines) and $a=\ln A$ (the solid lines) at different values of
  the parameter $\beta$: 1.5, 1.0 and 0.5, from top to bottom. Asymptotically
 $e^{2F}\rightarrow \frac{2\beta}{3}x^2$ and $A\rightarrow \frac{2\beta}{3}x^2$.}
\label{F_A_sphera}
  \end{center}
\end{minipage}\hfill
\end{figure}

In this subsection, we study the stability of the system of equations
\eqref{ein_wh_sine}-\eqref{field_wh_sine_0} when the radius of the wormhole throat goes to zero
$x_0 \rightarrow 0$, i.e., the spherically symmetric case. The static solutions for this case were found in
\cite{Dzhunushaliev:2009yw} as a limiting case of the wormhole solutions when $x_0\to 0$.
In this case, the only free parameter is $\beta$. We specify the boundary conditions as in
\cite{Dzhunushaliev:2009yw}
\begin{equation}
\label{ini_sphera_sine}
	A_0(0)=1, \quad \phi_0(0)={\rm const.}, \quad \phi_0^\prime(0)=0, \quad F_0(0)={\rm const.}
\end{equation}
With these initial conditions we numerically found solutions for $\phi (x), F(x), A(x)$
for different values of the parameter $\beta$ -- see figures \ref{phi_sphera}, \ref{F_A_sphera}.
The boundary conditions from \eqref{ini_sphera_sine} also imply we are focusing on the case
where the charges are set to zero since in the presence of charges the solutions become singular at $x\to 0$.
This can be seen from the boundary condition for $A (x)$ implied by equation \eqref{ein_wh_sine}:
 $$
 A_0(0):=1-\frac{\beta}{2}\frac{\bar{q}^2+\bar{Q}^2}{x^2}\to -\infty.
 $$

Some further analytical details of the linear stability analysis for the spherically symmetric case are given in appendix B.
The summary of the results from appendix B are that for the spherically symmetric ($x_0 = 0$) case,
the solutions are stable for $\beta > 3/4$ and unstable for $\beta < 3/4$.

\section{Conclusion}

In the paper we have studied dynamical stability of spherically symmetric and wormhole configurations for the
system of 4D gravity coupled to a sine-Gordon, ghost scalar field. In some we included cases electric and/or
magnetic charges. The stability of these solutions was tested with respect to linear, radial, harmonic
perturbations. For the wormhole configurations we considered the case both with and without electric and/or magnetic charges.
For the spherically symmetric case we investigated only the case when the electric and magnetic charges were zero, since
adding charges to the spherically symmetric solution would result in a singularity at $x=0$.

To determine whether a particular configuration was stable or not the equation for the perturbation of the scalar
field, \eqref{sf_pert}, was written in a Schr\"{o}dinger-like form \eqref{sf_pert_2}. The ``energy" eigenvalue of this
equation was the square of the frequency, $\omega ^2$, of the harmonic perturbations. If $\omega^2 >0$ then the solution
was stable; if $\omega^2 < 0$ this implied exponential growth of the perturbation and that the solution was unstable.
This method of testing for stability was used in \cite{jetzer, jetzer2} to study the stability of boson stars, and was
used by the present authors \cite{Dzh:2008} to study the stability of phantom/ghost boson stars. Other, stronger stability criteria
-- such as topological stability \cite{rajaraman} or the global stability of \cite{kusmartsev} -- were not applicable to
the present solution due to the asymptotic AdS nature of the space-time, the  asymptotic nature of the scalar field,
or the real character of the scalar field.

For the wormhole solutions, using a combination of numerical calculations and analytical estimates, it was shown that
stable solutions exist for a wide range of the parameters (e.g. the size of the wormhole throat $x_0$ and the free parameter $\beta$)
both with and without the charges. For the spherically symmetric solution (when the wormhole throat radius $x_0=0$),
it is shown that the stability of the system depends on the value of the parameter $\beta$: when
$\beta > 3/4$ the solutions were stable; when $\beta < 3/4$ the solutions were unstable.


The stability of these solutions, especially the spherically symmetric solutions, is a non-trivial result. Previously
the authors \cite{Dzh:2008} found spherically symmetric solutions to the system of 4D gravity plus a ghost scalar field having
a $\lambda \phi ^4$ self interaction. These solutions were compared to the Bartnik-McKinnon solutions \cite{bk} with the ghost field
playing the role of the SU(2) Yang-Mills fields.  However these ghost field supported solutions were not
stable. Here we find stable solutions by considering a ghost field with a sine-Gordon potential. In $(1+1)$ there exists the
well known sine-Gordon soliton \cite{rajaraman} but it is known from Derrick's theorem \cite{derrick} that soliton solutions
coming from only scalar fields do not exist in dimensions higher than $(1+1)$. What we have found here is a soliton-like solution in $(3+1)$
dimensions. This was achieved by adding gravity and letting the scalar field be a ghost field. The only feature about the spherical solutions
that is slightly undesirable is that fact that asymptotically the solution goes to AdS space-time rather than Minkowski
space-time. However, in light of the AdS/CFT correspondence \cite{Maldacena:1997re} and the connection between AdS space-time and
the holography ideas \cite{Witten:1998qj}, this feature of the asymptotic form of the solution may not be so negative. Finally,
if the universe has some cosmological constant as observations indicate, having a space-time with asymptotic dS or AdS
behavior is actually preferable to having asymptotic Minkowski behavior.

\begin{center}
\bf{Acknowledgment}
\end{center}

V.D. and V.F. are grateful to the Research Group Linkage
Programme of the Alexander von Humboldt Foundation for the
support of this research. They also would like to thank the
Carl von Ossietzky University of Oldenburg for hospitality
while this work was carried out.

\begin{center}
\bf{Appendix A: Details of Wormhole Case ($x_0 \ne 0$)}
\end{center}

In this appendix we give additional, analytical support for the positiveness
of the potential \eqref{pot_pert_2} (and therefore of $\omega ^2$)
by analyzing the asymptotic behavior of $V [x(\rho)]$ for $x \rightarrow 0$ and for $x \rightarrow \infty$.
First we consider the behavior of the potential \eqref{pot_pert_2} at $x \rightarrow 0$, making use
of the boundary conditions \eqref{ini_wh_sine}. To simplify the analysis we express the condition for
the metric function $A$ (i.e. the middle equation in \eqref{ini_wh_sine}) in term of $a=\ln{A}$. This yields
\begin{equation}
\label{ini_a0}
a_0(0)=\ln{\left(1-\beta\, \delta\, x_0^2\right)},
\end{equation}
where the index 0 refers to a static solution, and we have introduced the notation
$$
\delta=\frac{\gamma}{x_0^4}-2, \quad \gamma=\frac{\bar{q}^2 +\bar{Q}^2}{2}>0~.
$$
It follows from \eqref{ini_a0} that $A(0) \ge 0$. By taking the limit $A(0)=0$ one finds a lower bound on $x_0$, which
we denote as the critical value, $x_{cr}$:
\begin{equation}
\label{x_cr}
x_0^2:=x_{cr}^2>\frac{1}{4}\left[\frac{-1+\sqrt{1+8\beta^2 \gamma}}{\beta}\right].
\end{equation}
Taking into account this critical value, the behavior of the potential
\eqref{pot_pert_2} near $x=0$ depends on whether the parameter $\delta$ is positive or negative. In the case when
$\delta<0$ its value falls in the range
$$
-2\leq \delta <0 ~.
$$
The lower bound, $\delta=-2$, corresponds to absence of any charges
($\gamma=0$). There are two limiting cases:

(i) $x_0 \ll 1$, $\beta\approx 1$: in this case
$$a_0(0)\approx 0, \;F_0(0)\approx 0, \;\phi_0\approx \phi_0(0)+\phi_0^{\prime}x$$
and, accordingly, the potential \eqref{pot_pert_2} will then be
$$V[0]\approx 1+\frac{1}{x_0^2} \gg 1.$$
In other words the potential will be positive and large near $x =0$.

(ii) $x_0 \gg 1$: in this case $a_0(0)\approx \ln\left(\beta |\delta| x_0^2\right)$ and the potential \eqref{pot_pert_2} will be
$$V[0]\approx |\delta|\left(\frac{1}{2}+\beta\right).$$
Again the potential is positive near $x=0$.

We now turn to the asymptotic behavior of the potential \eqref{pot_pert_2} as $x \rightarrow \infty$.
In this case, the metric functions, $F(x) , A(x)$ behave as (see the discussion at the end
of section II and the caption for figure \ref{static_sols_2}):
$$e^{2F_0}\rightarrow \frac{2}{3}x^2, \quad a_0\rightarrow \ln{\left(\frac{2}{3} x^2\right)}.$$
To investigate the $x \rightarrow \infty$ behavior of the scalar field $\phi$ we expand the field as a
small perturbation, $\delta\phi_{\infty}$, around the asymptotic solution of $\pi$
$$
\phi=\pi+\delta\phi_{\infty}, \quad \delta\phi_{\infty} \ll 1.
$$
Plugging this form into equation \eqref{field_wh_sine_0} yields the following equation for $\delta\phi_{\infty}$
$$
\delta\phi_{\infty}^{\prime\prime}+\frac{4}{x}\delta\phi_{\infty}^{\prime}+\frac{3}{2x^2}\delta\phi_{\infty}=0.
$$
This equation has the solution
$$
\delta\phi_{\infty}=C x^{(-3+\sqrt{3})/2},
$$
where $C$ is an integration constant. Taking into account these asymptotic forms for the metric functions and the scalar field,
the behavior of the potential \eqref{pot_pert_2} is
$$
V[x(\rho)]_{\infty}\approx \frac{2}{9}x^2.
$$

In summary for the case $\delta<0$ we have the following behavior of the potential
from \eqref{pot_pert_2} near $x=0$ and as $x \rightarrow \infty$

\begin{tabular}{ll}
$x\to 0$: $V[x(\rho)]\to$  &  $\left\{  \begin{tabular}{l}
$\left(1+\frac{1}{x_0^2}\right) \gg 1$ \; at $x_0\ll 1, \; \beta\approx 1$;\\[\medskipamount]
$|\delta|\left(\frac{1}{2}+\beta\right) >0 $\; at $x_0\gg 1$;\\
\end{tabular}  \right.  $
\\[\bigskipamount] \\
$x\to \infty$: $V[x(\rho)]\to$&\,\,\, $\frac{2}{9}x^2,$   \; $x_0, \beta$ are arbitrary.\\[\medskipamount]
\end{tabular}

For the case $\delta >0 $, there is an upper limit, $\delta_{cr}$, coming from \eqref{x_cr}.
In particular, at $\gamma\to \infty$, $\delta_{cr}\to 0$, and at $\gamma\to 0$, $\delta_{cr}\to 1/(\beta^2 \gamma) \to \infty$.
Carrying out an analysis similar to the $\delta <0$ case, we find the following asymptotic behavior
for the potential from \eqref{pot_pert_2} for $0\leq\delta < \delta_{cr}$:

\begin{tabular}{ll}
$x\to 0$: $V[x(\rho)]\to$  &  $\left\{  \begin{tabular}{l}
$\left(1+\frac{1}{x_0^2}\right) >0$ \; at $\delta \to 0$, $x_0 > x_{cr}$ and finite;\\[\medskipamount]
$1$\; at $\delta \to \delta_{cr}, \, x_0 \to x_{cr}$;\\
\end{tabular}  \right.  $
\\[\bigskipamount] \\
$x\to \infty$: $V[x(\rho)]\to$&\,\,\, $\frac{2}{9}x^2,$   \; $x_0>x_{cr}$.\\[\medskipamount]
\end{tabular}

The above analysis shows that near $x=0$ and at $x\to \infty$ the potential
\eqref{pot_pert_2} is positive. Although we have no similar demonstration for the positiveness of
the potential at intermediate $x$, the numerical solutions of $V[x(\rho)]$ show that for some of the
parameters we investigated the potential was positive over the entire range of $x$ -- see figures \ref{stab_q_0},
\ref{omega_x0}. This shows that at least for these parameters the solutions found in section II and
in \cite{Dzhunushaliev:2009yw} are stable.

\begin{center}
\bf{Appendix B: Details of Spherical Symmetric Case ($x_0 =0$)}
\end{center}

As for the wormhole solutions, we now study the stability of the static solutions by looking at the behavior of the potential
\eqref{pot_pert_2} near $x\to 0$ and $x \to \infty$. First we expand the functions $\phi_0, A_0, F_0$
in a Taylor series in the neighborhood of $x=0$
\begin{eqnarray}
\label{ini_phi2}
\phi_0\approx \phi_0(0)+\phi_2\frac{x^2}{2},\\
\label{ini_A2}
A_0 \approx A_0(0)+A_2\frac{x^2}{2},\\
\label{ini_F2}
F_0 \approx F_0(0)+F_2\frac{x^2}{2},
\end{eqnarray}
where $\phi_2, A_2, F_2$ are the corresponding values of the second derivatives at
$x=0$. Expressions for these second derivatives can be found by inserting \eqref{ini_phi2}-\eqref{ini_F2}
into the static equations \eqref{ein_wh_sine}-\eqref{field_wh_sine_0}. This yields
$$
\phi_2=\frac{1}{3}\sin{\phi(0)}, \quad A_2=\beta\left[1-\cos{\phi(0)}\right], \quad F_2=\frac{1}{2}A_2.
$$
Inserting the expansions from \eqref{ini_phi2}-\eqref{ini_F2} into the potential \eqref{pot_pert_2}, gives
\begin{equation}
\label{V0_sphera}
V[0]= e^{2 F_0 (0)}\left[\beta+(1-\beta)\cos{\phi(0)} \right].
\end{equation}

\begin{figure}[h]
\begin{center}
  \includegraphics[width=11cm]{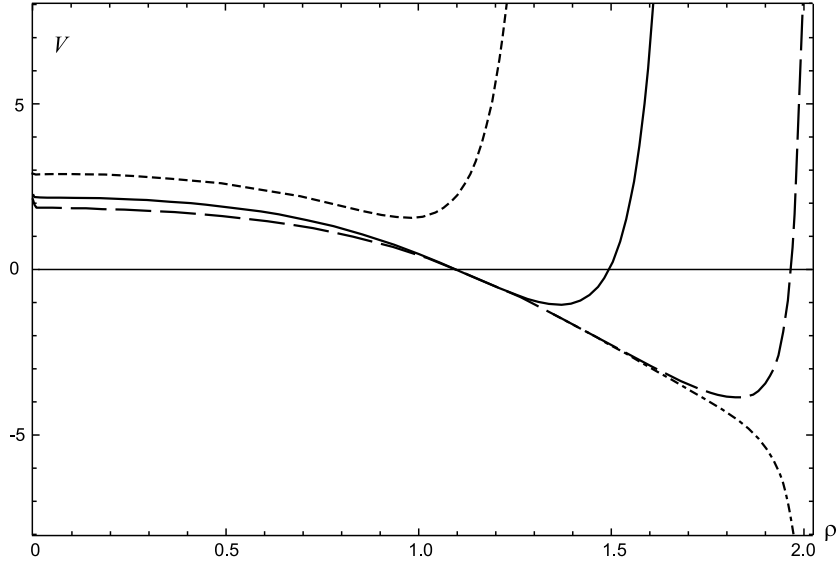}
 \caption{The potential $V[x(\rho)]$ from \eqref{pot_pert_2} as a function of $\rho$, where $\rho$ is defined in \eqref{new_indep}.
The the values $\beta= 1.5, 1.0, 0.76, 0.74$ correspond respectively to the short dashed line, the solid line, the long dashed line,
and the dash-dotted line. The critical value corresponds to $\beta=0.75$, see Eq. \eqref{Vinf_sphera}.}
\label{stab_sphera}
\end{center}
\end{figure}

We now turn to the $x\to \infty$ behavior of the potential \eqref{pot_pert_2}.
In this case, the metric functions behave as
$$
A_0\to \frac{2\beta}{3}x^2, \quad e^{2F_0}\to\frac{2\beta}{3}x^2.
$$
As for the wormhole solutions, in order to study the behavior of the scalar field $\phi$ at infinity,
we take a perturbation around the asymptotic solution of the static equation
\eqref{field_wh_sine_0} of the form
$$
\phi=\pi+\delta\phi_{\infty}, \quad \delta\phi_{\infty} \ll 1.
$$
With this equation \eqref{field_wh_sine_0} takes the form
$$
\delta\phi_{\infty}^{\prime\prime}+\frac{4}{x}\delta\phi_{\infty}^{\prime}+\frac{3}{2\beta x^2}\delta\phi_{\infty}=0 ~.
$$
If $\beta\geq \frac{2}{3}$ this has the following solution
$$
\delta\phi_{\infty}=C x^{n}, \quad n=\frac{-3\sqrt{\beta}+\sqrt{9\beta-6}}{2\sqrt{\beta}} ~,
$$
where $C$ is an integration constant. If $\beta\leq \frac{2}{3}$ the solution is
$$
\delta\phi_{\infty}=C_1\frac{\sin\left(\frac{\sqrt{6-9\beta}}{2\sqrt{\beta}}\ln{x}\right)}{x^{3/2}} +
C_2\frac{\cos\left(\frac{\sqrt{6-9\beta}}{2\sqrt{\beta}}\ln{x}\right)}{x^{3/2}}
,
$$
where $C_1, C_2$ are integration constants.

\begin{figure}[t]
\begin{center}
  \includegraphics[width=9cm]{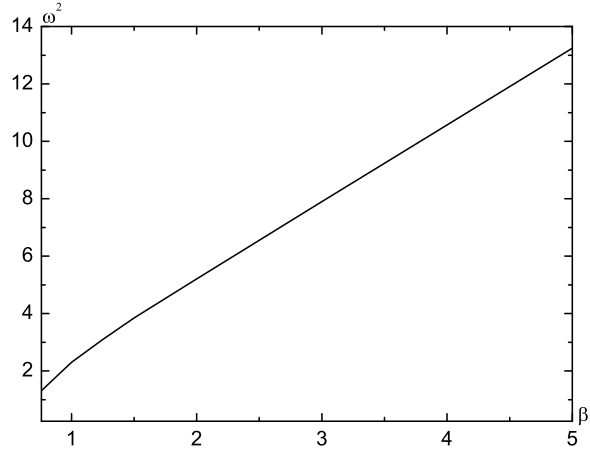}
\vspace{-1cm}
 \caption{The dependence of the lowest eigenvalues $\omega^2$ on the parameter $\beta$.}
\label{omega_sphera}
\end{center}
\end{figure}

Using these asymptotic expressions for the metric functions and the scalar field,
we can obtain the asymptotic behavior of the potential
\eqref{pot_pert_2} for the two cases:

(i) For $\beta\geq 2/3$:
$$
V[x(\rho)]_{\infty}=\frac{4}{9}\beta^2 x^2\left[2-\frac{3}{2\beta}+\frac{3}{2}n C^2 x^{2n}\right],
$$
where
\begin{center}
\begin{tabular}{ll}
$-1.5\leq n < 0$ \,\, since \quad &  $\left\{  \begin{tabular}{l}
$\delta\phi_{\infty}\to C x^{-1.5}$ at $\beta \to \frac{2}{3}$;\\[\medskipamount]
$\delta\phi_{\infty}\to C x^{-1/(2\beta)}$ at $\beta \to \infty$.\\
\end{tabular}  \right.  $
\end{tabular}
\end{center}

(ii) For $\beta\leq 2/3$:
$$
V[x(\rho)]_{\infty}=\frac{4}{9}\beta^2 x^2\left[2-\frac{3}{2\beta}+\frac{[\cos, \sin]}{x}\right],
$$
where the notation $[\cos, \sin]$ denotes a set of terms containing the functions
$$
\cos\left(\frac{\sqrt{6-9\beta}}{2\sqrt{\beta}}\ln{x}\right) \quad \text{and} \quad
\sin\left(\frac{\sqrt{6-9\beta}}{2\sqrt{\beta}}\ln{x}\right).
$$

Hence, one can see that, asymptotically,  the terms proportional to  $x^{2n}$ in the first case and $1/x$
in the second case can be neglected. The final expression for the asymptotic form of
the potential for both cases ($\beta \le 2/3$ and $\beta \ge 2/3$) is
\begin{equation}
\label{Vinf_sphera}
V[x(\rho)]_{\infty}\approx \frac{4}{9}\beta^2 x^2\left[2-\frac{3}{2\beta}\right].
\end{equation}
It is clear from this expression that there is critical value of
$\beta=3/4$. For $\beta <3/4$ the potential becomes negative and the static solutions
of the system \eqref{ein_wh_sine}-\eqref{field_wh_sine_0} are unstable.

A graphic illustration of the behavior of the potential \eqref{pot_pert_2} at various values of
$\beta$ is presented in figure \ref{stab_sphera}, and the behavior of the lowest
eigenvalues $\omega^2$ as the dependence on $\beta$ is shown in figure \ref{omega_sphera}.

\end{document}